\documentclass[twocolumn,superscriptaddress,amssymb,amsmath,aps,longbibliography,nofootinbib,prl,floatfix]{revtex4-2}
\usepackage{graphicx}% Include figure files
\usepackage{dcolumn}% Align table columns on decimal point
\usepackage{bm}% bold math
\usepackage{float}
\usepackage{comment}
\usepackage{amsmath}
\usepackage{amssymb}
\usepackage{color}
\usepackage{subeqnarray}
\usepackage[utf8]{inputenc}
\usepackage[english]{babel}
\usepackage{mathrsfs}
\usepackage[bb=boondox]{mathalfa}
\usepackage{xcolor} 

\usepackage{amsbsy}
\usepackage{latexsym,epsfig,graphicx}
\usepackage{subfigure}
\usepackage{amsfonts}
\usepackage{amsmath}
\usepackage{xspace}
\usepackage{epstopdf}

\usepackage{array,booktabs}
\usepackage{babel}
\usepackage{multirow}
\usepackage[colorlinks=true, pdfstartview=FitV, linkcolor=red, citecolor=blue, urlcolor=blue]{hyperref}
\usepackage[normalem]{ulem}

\usepackage{cleveref,braket,xcolor}

\pdfoutput=1

\begin{document}

\title{Symmetry Protected Two-Photon Coherence Time}

\author{Xuanying Lai}
\affiliation{Department of Physics, The University of Texas at Dallas, Richardson, Texas 75080, USA}
\affiliation{Elmore Family School of Electrical and Computer Engineering, Purdue University, West Lafayette, Indiana 47907, USA}
\affiliation{Department of Physics and Astronomy, Purdue University, West Lafayette, Indiana 47907, USA}

\author{Christopher Li}
\affiliation{Department of Physics, The University of Texas at Dallas, Richardson, Texas 75080, USA}
\affiliation{Elmore Family School of Electrical and Computer Engineering, Purdue University, West Lafayette, Indiana 47907, USA}
\affiliation{Department of Physics and Astronomy, Purdue University, West Lafayette, Indiana 47907, USA}

\author{Alan Zanders}
\affiliation{Department of Physics, The University of Texas at Dallas, Richardson, Texas 75080, USA}

\author{Yefeng Mei}
\email{yefeng.mei@wsu.edu}
\affiliation{Department of Physics and Astronomy, Washington State University, Pullman, Washington 99164, USA}

\author{Shengwang Du}
\email{du350@purdue.edu}
\affiliation{Department of Physics, The University of Texas at Dallas, Richardson, Texas 75080, USA}
\affiliation{Elmore Family School of Electrical and Computer Engineering, Purdue University, West Lafayette, Indiana 47907, USA}
\affiliation{Department of Physics and Astronomy, Purdue University, West Lafayette, Indiana 47907, USA}

\date{\today}

\begin{abstract}
We report the observation of symmetry protected two-photon coherence time of biphotons generated from backward spontaneous four-wave mixing in laser-cooled $^{87}$Rb atoms. When biphotons are nondegenerate, non-symmetric photonic absorption loss results in exponential decay of the temporal waveform of the two-photon joint probability amplitude, leading to shortened coherence time. In contrast, in the case of degenerate biphotons, when both paired photons propagate with the same group velocity and absorption coefficient, the two-photon coherence time, protected by space-time symmetry, remains unaffected by medium absorptive losses. Our experimental results validate these theoretical predictions. This outcome highlights the pivotal role of symmetry in manipulating and controlling photonic quantum states.
\end{abstract}

\maketitle

\begin{figure*}[t]
\centering
\includegraphics[width=0.85\textwidth]{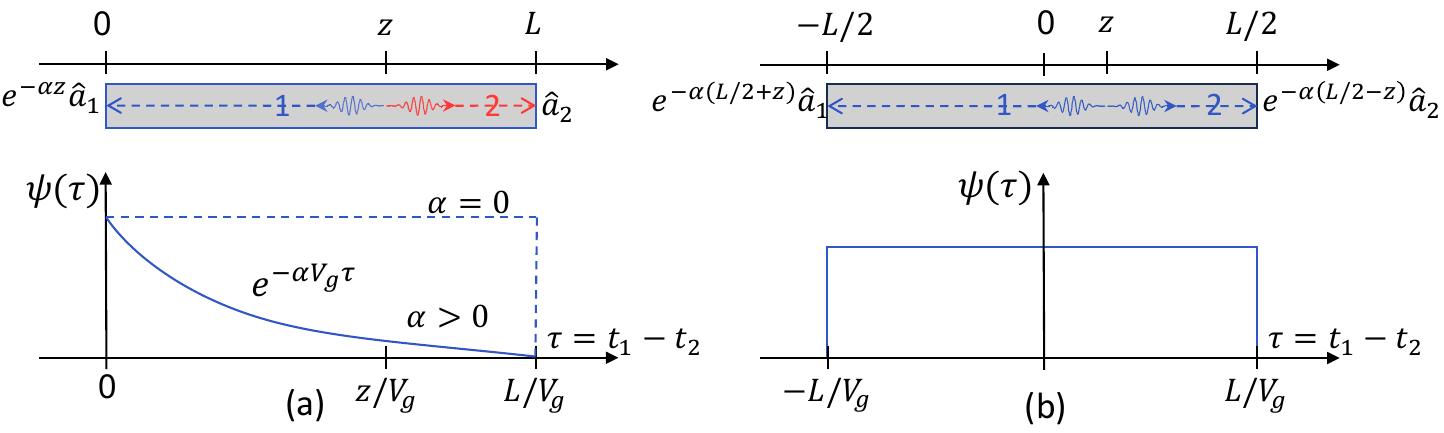}
\caption{Biphoton generation in absorptive media. (a) Nondegenerate biphoton generation: Photon 1 propagates along $-z$ direction with a slow group velocity $V_g\ll c$ and an absorption coefficient $\alpha$, while photon 2 propagates along $+z$ direction with the speed of light in vacuum $c$ and without any absorption loss. (b) Degenerate biphoton generation where both photons 1 and 2 counter propagate with the same slow group velocity $V_g$ and the same absorption coefficient $\alpha$.}
\label{fig:fig01}
\end{figure*}

Entangled photon pairs, termed \textit{biphotons}, generated from spontaneous parametric down conversion (SPDC) \cite{PhysRevLett.18.732, PhysRevLett.25.84,  PhysRevA.50.5122, YanhuaShih_2003} or spontaneous four-wave mixing (SFWM) \cite{SFWM_PhysRevA.70.031802, SFWM_Li:04, SFWM_Du:08, SFWM_Zhao:14, SFWM_PhysRevA.107.053703} have become benchmark tools in quantum optics, particularly for Bell inequality tests \cite{Bell_PhysRevLett.28.938, Bell_PhysRevLett.47.460, Bell_PhysRevLett.49.1804, Bell_PhysRevLett.115.250401}, two-photon interference \cite{TPI_PhysRevLett.59.2044, TWI_PhysRevA.85.021803}, quantum key distribution \cite{QKD_PhysRevLett.67.661, QKD_PhysRevA.76.012307}, and quantum teleportation \cite{QT_PhysRevLett.70.1895, QT_PhysRevLett.80.1121, QT_Nature1997}. Narrowband biphotons with long coherence times are crucial components for quantum networks \cite{KimbleQN} due to their efficient interaction with matter nodes, such as atoms and ions. Remarkably, in the absence of significant loss, the two-photon coherence time of biphotons generated in the phase-matching regime is determined by the relative group delay between the paired photons \cite{PhysRevA.50.5122, SFWM_Du:08, PhysRevLett.100.183603}, making it possible to control their temporal characteristics through material dispersion engineering \cite{PhysRevLett.100.183603}.  

In open photonic quantum systems, coherence time is typically shortened due to irreversible coupling with the environment, such as loss and dephasing \cite{RevModPhys.75.715, RevModPhys.76.1267}. This phenomenon also holds true for biphoton generation when the nonlinear medium (SPDC or SFWM) exhibits photon absorption. This physics is illustrated in Fig.~\ref{fig:fig01}(a). Consider nondegenerate backward biphotons generated within a uniform nonlinear medium of length $L$. Photon 1 propagates in the $-z$ direction with a slow group velocity $V_g \ll c$ and an absorption coefficient $\alpha$, while photon 2 moves in the $+z$ direction at the speed of light in vacuum $c$ with no absorption loss. A photon pair can be generated at any point with equal probability within the medium. Two single-photon counters are positioned at both surfaces ($z=0$ and $z=L$). For two paired photons generated at position $z$, they arrive at $z=0$ and $z=L$ with a relative time delay $\tau=t_1-t_2 \approx z/V_g$. In the absence of loss ($\alpha=0$), the two-photon coincidence joint probability amplitude appears as a rectangular waveform (dashed line in Fig.~\ref{fig:fig01}(a)), with a coherence time of $L/V_g$, determined by the relative group delay. However, in the presence of absorption ($\alpha > 0$), the field of photon 1 at the output surface $z=0$ is attenuated to $e^{-\alpha z}\hat{a}_1,$ while photon 2 ($\hat{a}_2$) remains unaffected. Consequently, the two-photon coincidence is registered with a relative time delay $\tau=z/V_g,$ yielding a two-photon joint probability amplitude
\begin{equation}
\begin{aligned}
\psi(\tau)=\langle0,0|e^{-\alpha z}\hat{a}_1\hat{a}_2|1,1\rangle\propto e^{-\alpha V_g \tau},
\end{aligned}
\label{eq:psi1}
\end{equation}
as depicted by the solid blue curve in Fig.~\ref{fig:fig01}(a). When the loss is significant, the resulting two-photon coherence time $1/(2\alpha V_g)$ is significantly shortened compared to $L/V_g$. This exponential decay waveform is consistent with the results obtained through theory in the interaction picture \cite{SFWM_Du:08}.

Now, let's consider degenerate biphoton generation, where both counter-propagating photons have the same frequency, as shown in Fig.~\ref{fig:fig01}(b). They propagate with the same group velocity $V_g$ and the same absorption coefficient $\alpha$. To count the symmetry, we choose the original position $z=0$ at the center of the medium. When a photon pair is generated at position $z$, the field of photon 1, propagating in the $-z$ direction and arriving at $z=-L/2$, becomes $e^{-\alpha(L/2+z)}\hat{a}_1$, while the field of photon 2, propagating in the $+z$ direction and arriving at $z=L/2$, is $e^{-\alpha(L/2-z)}\hat{a}_2$. Thus the two-photon joint probability amplitude, $e^{-\alpha L}$, is independent of position $z$. As the relative time delay at the two output surfaces is $\tau=2z/V_g$, the two-photon joint probability amplitude retains its rectangular shape
\begin{equation}
\begin{aligned}
\psi(\tau)=\langle0,0|e^{-\alpha L}\hat{a}_1\hat{a}_2|1,1\rangle\propto e^{-\alpha L}\sqcap(\tau; -L/V_g, L/V_g).
\end{aligned}
\label{eq:psi2}
\end{equation}
The rectangular function $\sqcap$, ranging from $\tau=-L/V_g$ to $L/V_g$, illustrates that the two-photon coherence time is extended to $2L/V_g$. Although the two-photon joint probability amplitude is reduced by a factor of $e^{-\alpha L}$, resulting in a lower coincidence rate, the coherence time of $2L/V_g$ is preserved and protected by symmetry. Unlike the nondegenerate case, symmetry-protected two-photon coherence is independent of photonic absorptive loss.

\begin{figure}[t]
\centering
\includegraphics[width=\linewidth]{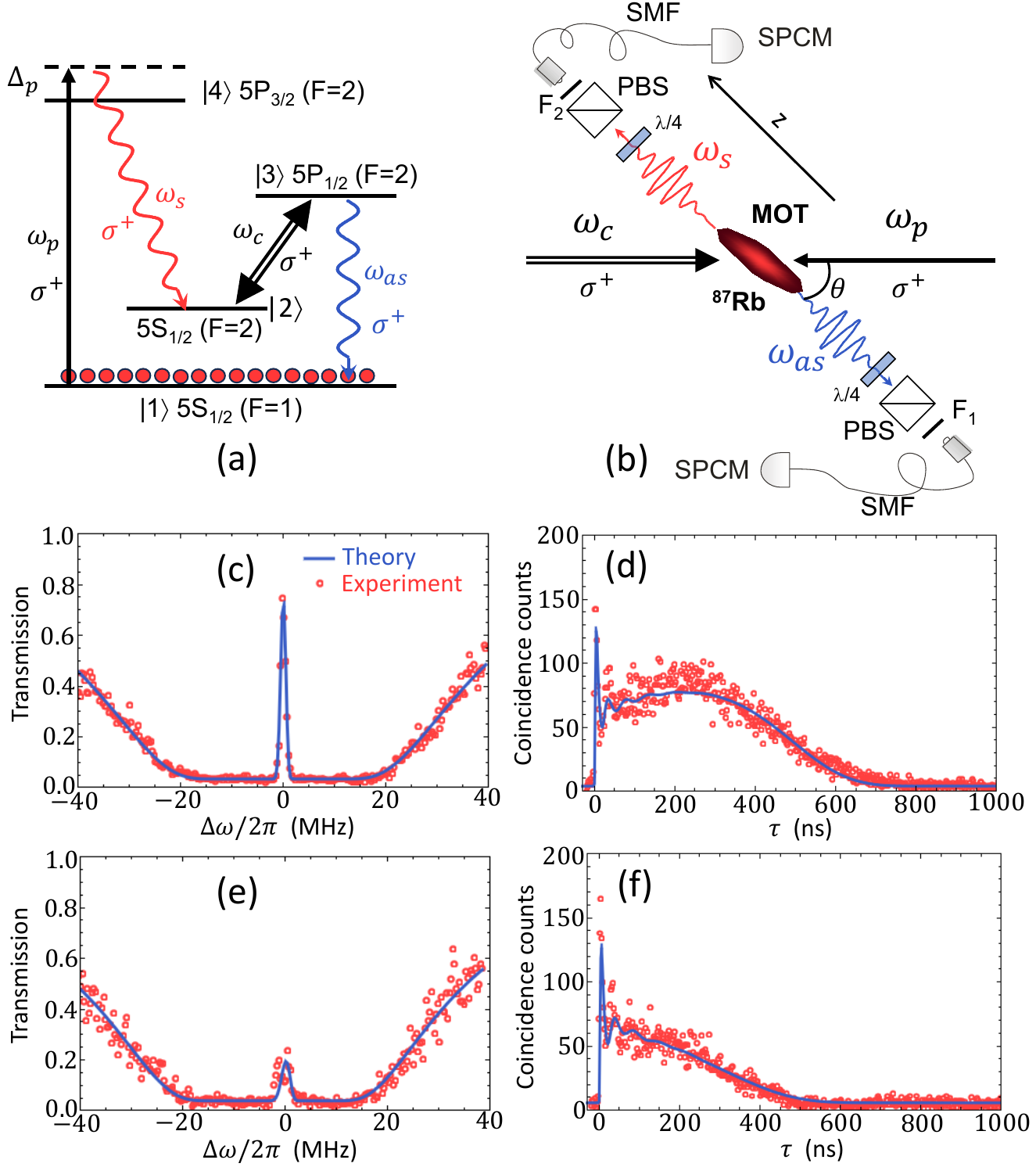}
\caption{Nondegenerate biphoton generation via SFWM in cold $^{87}$Rb atoms. (a) $^{87}$Rb atomic energy level diagram. (b) SFWM optical setup. (c) EIT spectrum with  $\alpha L =0.013$ on resonance. (d) Biphoton coincidence counts under the EIT condition of (c). (e) EIT spectrum with $\alpha L =0.71$ on resonance. (f) Biphoton coincidence counts under the EIT condition of (e). The coincidence collection time is $T_c=10$ min. The time binwidth is 2 ns.}
\label{fig:NondegenerateBiphoton}
\end{figure}
To experimentally validate the aforementioned predictions, we initially generate nondegenerate narrowband biphotons from SFWM \cite{SFWM_Du:08, PhysRevLett.100.183603, SFWM_Zhao:14} in a cloud of $^{87}$Rb atoms confined in a two-dimensional (2D) magneto-optical trap (MOT) with a temperature of about 90 $\mu$K \cite{Zhang2012}, as illustrated in Fig.~\ref{fig:NondegenerateBiphoton}. The length of the atomic cloud is $L = 1.7$ cm along its longitudinal $z$ direction. The experiment operates periodically with a cycle duration of 2.5 ms, including a MOT loading and state preparation time of 2.4 ms and a biphoton generation time of 0.1 ms. After the MOT loading time, the atoms are optically pumped to the ground state $|1\rangle$, as depicted in the energy level diagram in Fig.~\ref{fig:NondegenerateBiphoton}(a). The SFWM process is driven by a pair of counter-propagating pump ($\omega_p$, D2 line, 780 nm) and coupling ($\omega_c$, D1 line, 795 nm) laser beams, aligned at an angle of $\theta=3^o$ to the longitudinal $z$ axis, as shown in Fig.~\ref{fig:NondegenerateBiphoton}(b). The circularly ($\sigma^+$) polarized pump laser beam with Rabi frequency $\Omega_p=2\pi\times3.6$ MHz is blue-detuned by $\Delta_p=2\pi\times200~\rm MHz$ from the transition $|1\rangle\rightarrow|4\rangle$. The coupling laser beam ($\sigma^+$, $\Omega_c=2\pi\times12.2$ MHz) is on resonance with the transition $|2\rangle\rightarrow|3\rangle$. Phase-matched backward nondegenerate Stokes ($\omega_s$, 780 nm) and anti-Stokes ($\omega_{as}$, 795 nm) photon pairs are spontaneously produced and collected by a pair of opposing single-mode fibers (SMFs) placed along the MOT longitudinal $z$ axis. We use two narrowband etalon filters F$_1$ and F$_2$ to filter away stray lights. Photon coincidence counts are recorded by two single-photon counting modules (SPCMs). The atomic optical depth (OD) in the anti-Stokes transition is 88 (See Supplemental Material \cite{SuppM} for more detailed experimental parameters). In this nondegenerate case, the anti-Stokes photons propagate with a slow group velocity due to the effect of electromagnetically induced transparency (EIT) \cite{EITPhysicsTodayHarris, EITRevModPhys}. The far-off resonance Stokes photons propagate at nearly the speed of light in vacuum $c$ with negligible absorption loss.

In our EIT system, the anti-Stokes group delay time can be estimated as $L/V_g \simeq (2\gamma_{13}/|\Omega_c|^2)\mathrm{OD}$, and its absorption loss follows $\alpha L=2 \mathrm{OD} \gamma_{12}\gamma_{13}/(|\Omega_c|^2+4\gamma_{12}\gamma_{13})$ \cite{SFWM_Du:08}. Here $\gamma_{12}$ is the dephasing rate between the two ground states $|1\rangle$ and $|2\rangle$, and $\gamma_{13}=2\pi\times 3$ MHz. To keep the group delay time unaffected, we can control the EIT transmission or absorption coefficient by tuning the ground-state dephasing rate $\gamma_{12}$. To achieve a small $\gamma_{12}$, we switch off the MOT magnetic field during the biphoton generation time and cancel the Earth's residual field with external bias coils. The optimized EIT transmission spectrum is shown in Fig.~\ref{fig:NondegenerateBiphoton}(c). The EIT resonance is nearly transparent with a transmission of 97\%, corresponding to a dephasing rate $\gamma_{12}=2\pi\times0.0042$ MHz and absorption $\alpha L =0.013$. The corresponding biphoton correlation $|\psi(\tau)|^2$, measured as two-photon coincidence counts, shown in Fig.~\ref{fig:NondegenerateBiphoton}(d), displays a rectangular-like waveform with a $e^{-1}$ coherence time of 555 ns, consistent with the group delay time obtained from the EIT measurement. The oscillatory peak in the leading edge is the biphoton optical precursor \cite{BiphotonPrecursorOLDu2008, SinglePhotonPrecursorPRLDu2011}. The Gaussian-like tail results from the spatial Gaussian profiles of the pump and coupling laser beams \cite{PhysRevLett.115.193601, SFWM_Zhao:14, PhysRevA.93.033815}. 

Subsequently, we apply an external magnetic field gradient along the $z$ axis to increase the dephasing rate to $\gamma_{12}=2\pi\times0.20$ MHz. The resulting EIT transmission spectrum is shown in Fig.~\ref{fig:NondegenerateBiphoton}(e), with the resonance transmission reduced to only 24\% ($<e^{-1}$), corresponding to $\alpha L =0.71$. As predicted, the biphoton correlation displays an exponential decay waveform in Fig.~\ref{fig:NondegenerateBiphoton}(f). The fitted exponential decay time constant is 340 ns, consistent with that obtained from $1/(2\alpha V_g)=400$ ns with $\alpha=41.8~\rm m^{-1}$ and $V_g=3.0\times 10^4$ m/s from the EIT measurement. As expected, the two-photon coherence time is shortened due to nonsymmetric absorption of the paired photons.

\begin{figure}[t]
\centering
\includegraphics[width=\linewidth]{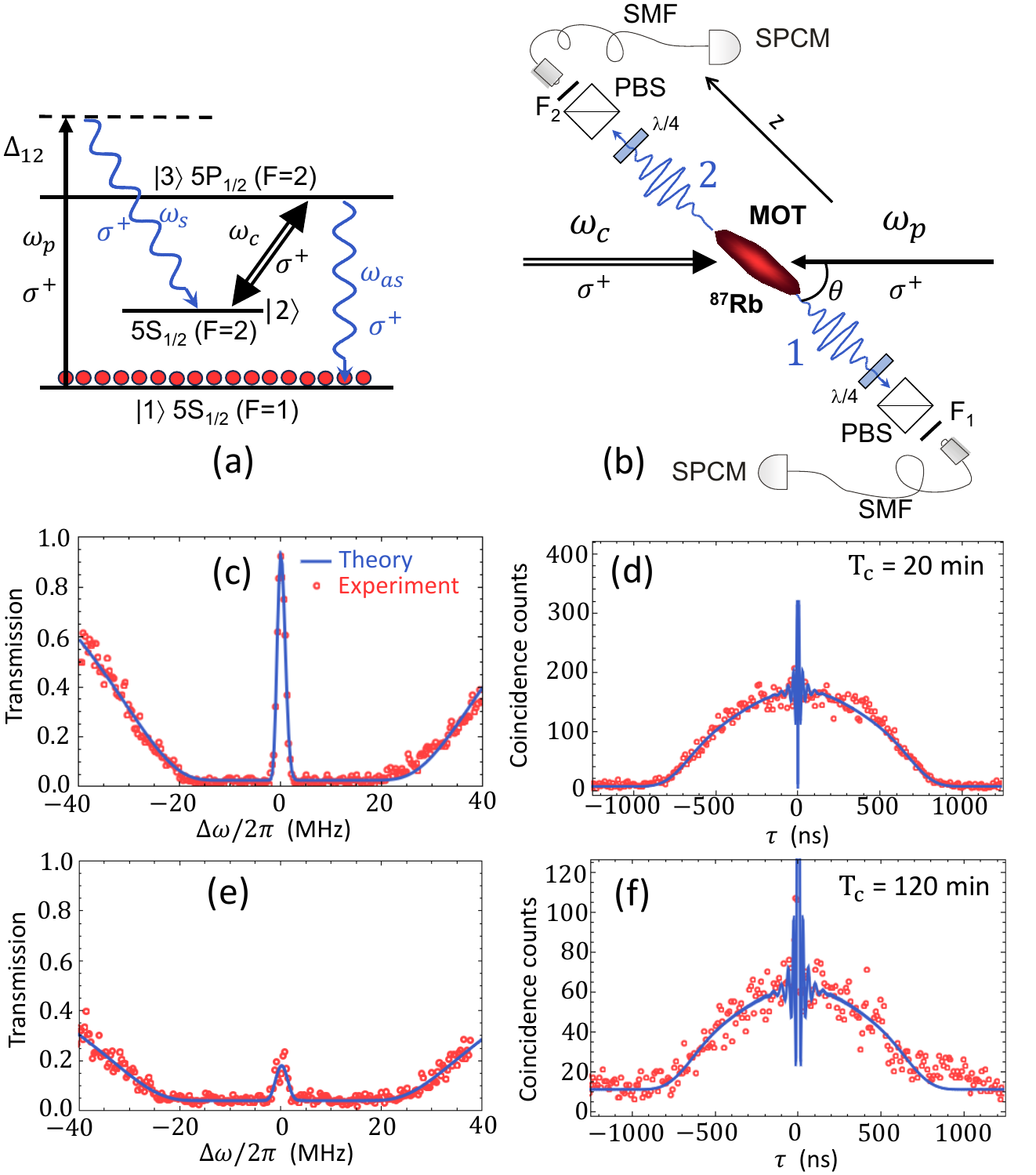}
\caption{Degenerate biphoton generation with symmetry protected two-photon coherence. (a) $^{87}$Rb atomic energy level diagram. (b) SFWM optical setup. (c) EIT spectrum with $\alpha L = 0.017$ on resonance. (d) Biphoton coincidence counts under the EIT condition of (c) with $\Omega_p=2\pi\times218.6$ MHz. (e) EIT spectrum with $\alpha L = 0.85$ on resonance. (f) Biphoton coincidence counts under the EIT condition of (e) with $\Omega_p=2\pi\times178.5$ MHz. The time binwidth is 10 ns.}
\label{fig:DegenerateBiphoton}
\end{figure}

We now proceed to experimentally demonstrate degenerate biphotons whose two-photon coherence time is protected by symmetry. As illustrated in Fig.~\ref{fig:DegenerateBiphoton}(a) and (b), the optical setup is nearly identical to that for nondegenerate biphoton generation, but with both pump and coupling lasers now in the $^{87}$Rb D1 line (795 nm) (See Supplemental Material \cite{SuppM} for more experimental details). The pump beam is blue-detuned from the transition $|1\rangle\rightarrow|3\rangle$ by $\Delta_{12}=2\pi\times6.8$ GHz, where $\Delta_{12}$ is the hyperfine splitting between the two ground levels $|1\rangle$ and $|2\rangle$. The coupling beam is still on resonance to the transition $|2\rangle\rightarrow|3\rangle$ with $\Omega_c=2\pi\times14.5$ MHz. In this configuration, both spontaneously generated Stokes and anti-Stokes photons are at the same frequency and with the same polarization. Consequently, they experience the same EIT group delay and absorption, fulfilling the symmetry described in Fig.~\ref{fig:fig01}(b). The OD for both photons is 150. Similar to the nondegenerate experiment, we vary the medium absorption by tuning the ground-state dephasing rate. Figure~\ref{fig:DegenerateBiphoton}(c) shows the EIT transmission with small absorption $\alpha L =0.017$. The corresponding biphoton correlation waveform is displayed in Fig.~\ref{fig:DegenerateBiphoton}(d). Although there is an oscillatory structure around $\tau=0$ (biphoton precursor \cite{BiphotonPrecursorOLDu2008, SinglePhotonPrecursorPRLDu2011}) and smoothly turning-off tails at the two ends resulting from the spatial Gaussian profiles of the pump and coupling laser beams \cite{PhysRevLett.115.193601}, the overall waveform exhibits a symmetric rectangular-like shape as expected. We now increase the medium absorption to $\alpha L =0.85$ and plot the EIT transmission in Fig.~\ref{fig:DegenerateBiphoton}(e). Unlike the nondegenerate case, the symmetric rectangular-like shape waveform is preserved, though the coincidence counts are reduced due to the larger absorption. The two-photon coherence time of  1,250 ns is consistent with the group delay time $2L/V_g$ =1350 ns obtained from the EIT measurements. Thus, we experimentally confirm the symmetry-protected two-photon coherence time in degenerate backward biphoton generation.

To verify the nonclassical property of the biphoton source, we confirm its violation of the Cauchy–Schwartz inequality \cite{PhysRevD.9.853}. Normalizing the coincidence counts to the accidental background floor in Fig.~\ref{fig:DegenerateBiphoton}(d), we obtain the normalized cross-correlation function $g^{(2)}_{12}(\tau)$ with a peak value of 30 $\pm$ 11. With the measured autocorrelations $g^{(2)}_{11}(0)=g^{(2)}_{11}(0)=2 \pm 0.2$, we get  $[g^{(2)}_{12}(\tau)]_{\max}^2/[g^{(2)}_{11}(0)g^{(2)}_{11}(0)]=$219 $\pm$ 171, violating the Cauchy–Schwartz inequality by a factor of 219.

These narrowband time-energy entangled biphotons can be used to produce heralded single photons with high purity \cite{PhysRevA.92.043836}, whose conditional autocorrelation function offers another quantitative measure of the quantum nature of the source \cite{PGrangier_1986}. Triggered by the detection of photon 1, we pass its paired photon 2 through a beam splitter (BS) whose outputs are connected to two SPCMs. For the biphotons in Fig.~\ref{fig:DegenerateBiphoton}(d), the measured conditional autocorrelation function is $g_{2|1}^{(2)}(0)=0.337 \pm 0.069$ with a time window of 1,200 ns, which is below the two-photon threshold value of 0.5.

\begin{figure}[t]
\centering
\includegraphics[width=0.95\linewidth]{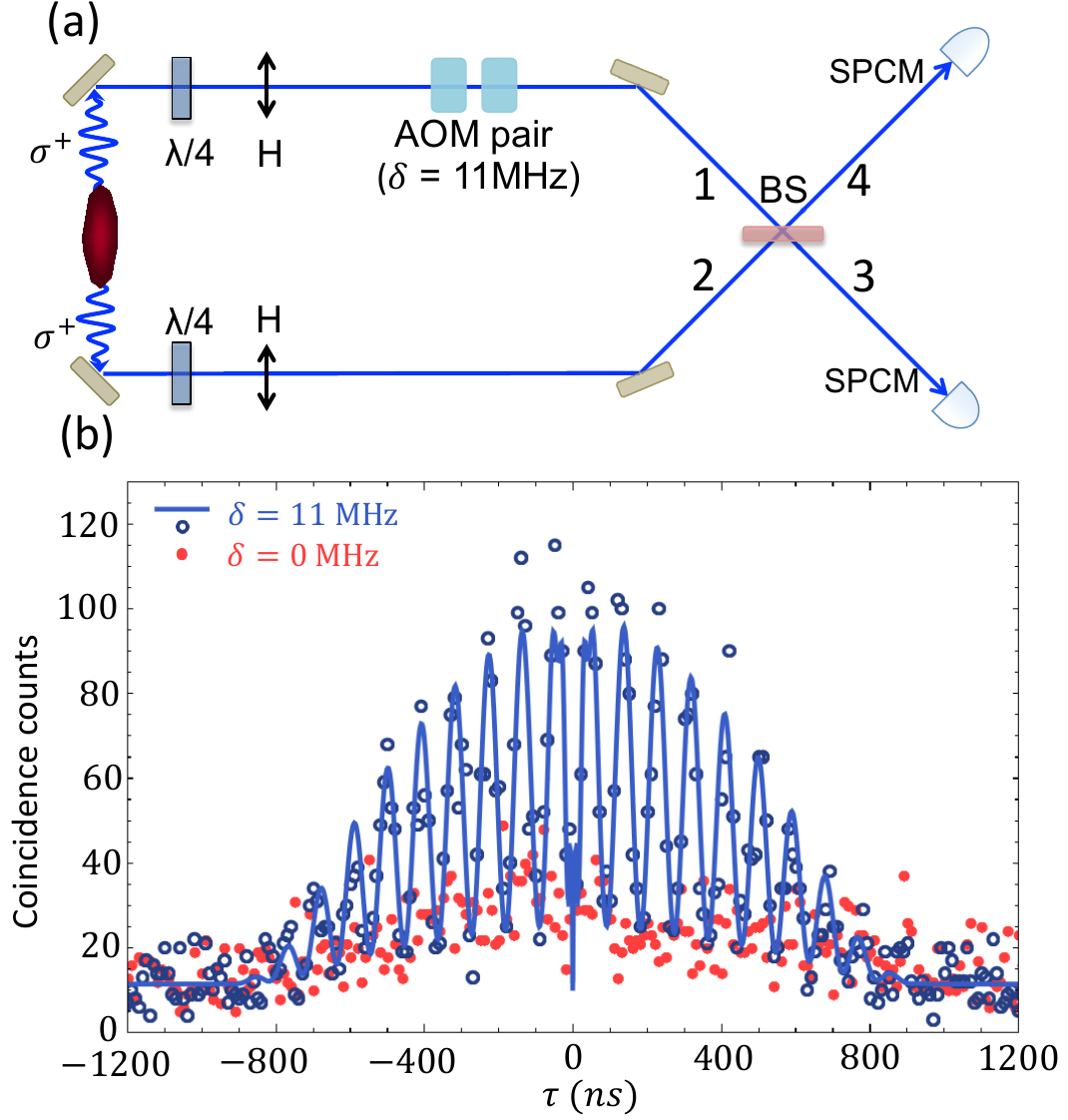}
\caption{Biphoton beating experiment. (a) Optical setup. (b) Biphoton beating measured as coincidence counts.  }
\label{fig:BiphotonBeating}
\end{figure}

We further confirm that the symmetry-protected two-photon correlation time preserves their phase coherence by measuring two-photon interference \cite{APB.77.797, TWI_PhysRevA.85.021803, TPI_PhysRevLett.59.2044}. The experimental setup for observing two-photon time-resolved interference is depicted in Fig.~\ref{fig:BiphotonBeating}(a). Through the use of quarter-wave ($\lambda/4$) plates, the $\sigma^+$ circular polarization of photons 1 and 2 is transformed into horizontal (H) linear polarization. Photon 1 is then subjected to an upper frequency shift of $\delta=$ 11 MHz by passing through a pair of acousto-optic modulators (AOMs). Subsequently, both photons are directed into a BS, with their outputs, labeled 3 and 4, detected by two SPCMs. The coincidence pattern, as shown in Fig.~\ref{fig:BiphotonBeating}(b), exhibits a two-photon beating oscillation at a frequency of 11 MHz, with a peak visibility of  $78 \pm 4\%$. The deviation from perfect visibility is attributed to the imbalance of the BS, which has a splitting ratio of $30\%:70\%$ in our experiment (See Supplemental Material \cite{SuppM}). These findings demonstrate that the correlation time, as measured in Fig.~\ref{fig:DegenerateBiphoton}(d), indeed corresponds to the phase coherence time of the biphoton joint amplitude.

\begin{figure}[t]
\centering
\includegraphics[width=0.9\linewidth]{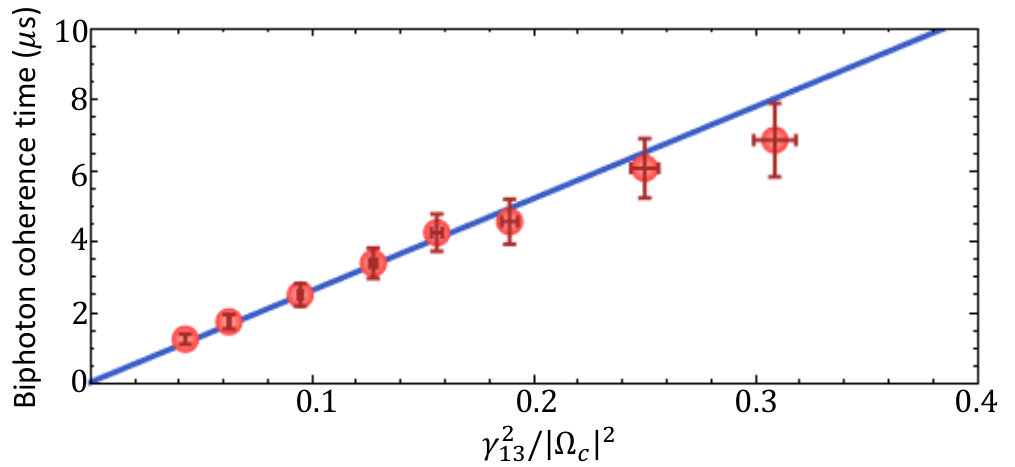}
\caption{Biphoton coherence time as a function of $\gamma_{13}^2/|\Omega_c|^2$. The solid line is calculated from $2L/V_g=(4\gamma_{13}/|\Omega_c|^2)\mathrm{OD}$, and the circles are experimental data. OD = 150.}
\label{fig:CoherencetimeVSCouplingPower}
\end{figure}

The EIT group delay time can be estimated as $L/V_g \simeq (2\gamma_{13}/|\Omega_c|^2)\mathrm{OD}$. If absorption loss is not present, we expect the two-photon coherence time to be determined by the group delay time for both nondegenerate and degenerate cases. Therefore, we can increase the two-photon coherence time by reducing the group velocity with a fixed medium length. In a perfect EIT medium with $\gamma_{12}=0$, this can be achieved by reducing the coupling laser power. However, in a realistic EIT medium with a finite dephasing rate $\gamma_{12}\neq 0$, the absorption loss, $\alpha L\simeq2 \mathrm{OD} \gamma_{12}\gamma_{13}/(|\Omega_c|^2+4\gamma_{12}\gamma_{13})$, increases as we reduce the coupling laser power. As a result, for the nondegenerate case, the biphoton coherence time cannot be further increased by reducing the coupling laser power as the absorption loss becomes significant. However, this problem can be overcome in the degenerate case, whose coherence time is protected by symmetry. We confirm this by measuring the two-photon coherence time as a function of $\gamma_{13}^2/|\Omega_c|^2$, as shown in Fig.~\ref{fig:CoherencetimeVSCouplingPower}, in which the coherence time increases from 1.25 $\mu$s to 6.85 $\mu$s, corresponding to the joint spectral bandwidth from 600 kHz to 80 kHz, by varying the coupling laser power.

The theoretical curves (solid lines in Fig.~\ref{fig:NondegenerateBiphoton}(c)-(f) and Fig.~\ref{fig:DegenerateBiphoton}(c)-(f)) are obtained through perturbation theory in the interaction picture \cite{SFWM_Du:08} (see Supplemental Material \cite{SuppM}). The theory aligns well with the experimental results. In the group delay regime, the biphoton wave function is the Fourier transform of its longitudinal detuning function $\Phi=\mathrm{sinc}(\Delta k L/2)e^{i(k_1+k_2)L/2}$ \cite{SFWM_Du:08}, where $\Delta k\simeq k_1-k_2-(k_c-k_p)\cos{\theta}$ is the complex phase mismatching \cite{SFWM_PhysRevA.107.053703}. Here, $\vec{k}_m$ are the wave vectors of the fields. In the nondegenerate case with significant loss, the longitudinal detuning function can be approximated as $\Phi\simeq i/(\Delta\omega L/V_g+i\alpha L)$ \cite{SFWM_Du:08}, whose Fourier transform gives an exponential decay waveform in Eq.~\eqref{eq:psi1}. In the degenerate case, the complex EIT wave numbers near resonance are $k_1\simeq k_0 + \Delta\omega/V_g + i\alpha$ and $k_2\simeq k_0 - \Delta\omega/V_g + i\alpha$, leading to the cancellation of loss in the phase mismatching $\Delta k\simeq 2\Delta\omega/V_g$. As a result, the longitudinal detuning function becomes
\begin{equation}
\begin{aligned}
\Phi(\Delta\omega)\simeq \mathrm{sinc}(\frac{\Delta\omega L}{V_g}\big{)} e^{-\alpha L},
\end{aligned}
\label{eq:ldf}
\end{equation}
whose Fourier transform is exactly the biphoton rectangular wave function of Eq.~\eqref{eq:psi2}. The symmetry cancels losses in the complex phase mismatching, thus protecting the two-photon coherence time.

We can also comprehend the symmetry in the Heisenberg picture. On EIT resonance, the field operators of counter-propagating degenerate biphotons are governed by the following coupled equations \cite{SFWM_PhysRevA.107.053703}:
\begin{equation}
\begin{aligned}
i\frac{\partial}{\partial z}\left[\begin{matrix}\hat a_{1}\\\hat a_2^\dagger\\\end{matrix}\right]
=\left[\begin{matrix}-i\alpha&-\kappa\\-\kappa&i\alpha\\\end{matrix}\right]\left[\begin{matrix}\hat a_{1}\\{\hat a_2^{\dagger}}\\\end{matrix}\right]+\left[\begin{matrix}\hat F_{1}\\{\hat F_2^{\dagger}}\\\end{matrix}\right],
\end{aligned}\label{eq:PTFWM}
\end{equation}
where $\kappa$ is the real nonlinear coupling coefficient, $\hat F_{1}$ and $\hat F_{2}$ are Langevin noise operators. Interestingly, the two-mode coupling matrix in Eq. \eqref{eq:PTFWM} follows parity-time (PT) symmetry \cite{PhysRevLett.80.5243, Miri_2019}. In conventional systems, balanced gain and loss give rise to their PT symmetry. In the backward degenerate biphoton generation, the PT symmetry effectively ``turns" the loss of one mode into ``gain" to compensate for the loss in another mode so that the coherence of the two-photon joint amplitude is protected.

The work was supported by DOE (DE-SC0022069), AFOSR (FA9550-22-1-0043) and NSF (2114076, 2228725). Y. M. acknowledges the startup funding from WSU.

\bibliography{biphoton}

\end{document}

\preprint{APS/123-QED}

\title{Supplemental Material for ''Symmetry Protected Two-Photon Coherence Time"}%

\author{Xuanying Lai}
\affiliation{Department of Physics, The University of Texas at Dallas, Richardson, Texas 75080, USA}
\affiliation{Elmore Family School of Electrical and Computer Engineering, Purdue University, West Lafayette, Indiana 47907, USA}
\affiliation{Department of Physics and Astronomy, Purdue University, West Lafayette, Indiana 47907, USA}

\author{Christopher Li}
\affiliation{Department of Physics, The University of Texas at Dallas, Richardson, Texas 75080, USA}
\affiliation{Elmore Family School of Electrical and Computer Engineering, Purdue University, West Lafayette, Indiana 47907, USA}
\affiliation{Department of Physics and Astronomy, Purdue University, West Lafayette, Indiana 47907, USA}

\author{Alan Zanders}
\affiliation{Department of Physics, The University of Texas at Dallas, Richardson, Texas 75080, USA}

\author{Yefeng Mei}
\email{yefeng.mei@wsu.edu}
\affiliation{Department of Physics and Astronomy, Washington State University, Pullman, Washington 99164, USA}

\author{Shengwang Du}
\email{du350@purdue.edu}
\affiliation{Department of Physics, The University of Texas at Dallas, Richardson, Texas 75080, USA}
\affiliation{Elmore Family School of Electrical and Computer Engineering, Purdue University, West Lafayette, Indiana 47907, USA}
\affiliation{Department of Physics and Astronomy, Purdue University, West Lafayette, Indiana 47907, USA}

%\date{\today}

\maketitle

\section{S1. Experimental parameters}

In this section, we present the parameters used for the experimental setup in the main manuscript.
~\\

\noindent\textbf{\textit{Overall Setting}}

%- Atomic temperature: 90 $\mu$K

- The longitudinal length of atomic cloud: $L = 1.7$ cm

- Intersection angle: $\theta = 3^o$

- Etalon filters ($F_1$  and $F_2$): 

\qquad Bandwidth 140 MHz

\qquad Isolation ratio 60 dB

- Duty cycle: $\eta_d=4\%$

- Total joint two-photon detection efficiency:

\qquad $\eta_c=\eta_f\eta_{t_1}\eta_{t_2}\eta_q^2=4.9\%$

\qquad Fiber coupling efficiency $\eta_f=70\%$

\qquad Optical transmission $\eta_{t_1}=48\%$ and $\eta_{t_2}=58\%$

\qquad Quantum efficiencies of two SPCMs $\eta_{q}=50\%$

- Rabi frequency: $\Omega=\frac{\mu E}{\hbar}=\sqrt{\frac{2I_0}{\epsilon_0 c}}\frac{\mu}{\hbar}$, where $I_0=\frac{2P}{\pi w_0^2}$ is the peak intensity of Gaussian profile, with electric dipole $\mu$, power $P$ and $e^{-2}$-intensity waist diameter $w_0$.
~\\

\noindent\textbf{\textit{Nondegenerate Biphoton Generation}}

- Relative energy levels:

\qquad $|1\rangle=|5S_{1/2}, F=1\rangle$

\qquad $|2\rangle=|5S_{1/2}, F=2\rangle$

\qquad $|3\rangle=|5P_{1/2}, F=2\rangle$

\qquad $|4\rangle=|5P_{3/2}, F=2\rangle$

- Pump beam:

\qquad Wavelength $\lambda_p$ = 780 nm

\qquad Detuning to $|1\rangle \rightarrow |4\rangle$, $\Delta_p=2\pi\times200~\rm MHz$

\qquad Polarization $\sigma^+$

\qquad Power 0.165 mW

\qquad $e^{-2}$-intensity waist diameter 2.9 mm

\qquad Rabi frequency $\Omega_p=2\pi\times3.6$ MHz

- Coupling beam:

\qquad Wavelength $\lambda_c$ = 795 nm

\qquad On resonance with $|2\rangle \rightarrow |3\rangle$

\qquad Polarization $\sigma^+$

\qquad Power 3.0 mW

\qquad $e^{-2}$-intensity waist diameter 2.9 mm

\qquad Rabi frequency $\Omega_c=2\pi\times12.2$ MHz

- Optical depth OD = 88

- Detection time $T_c$ = 10 min

- Time bin width $\Delta t_{bin}$ = 2 ns
~\\

\noindent\textbf{\textit{Degenerate Biphoton Generation}}

- Relative energy levels:

\qquad $|1\rangle=|5S_{1/2}, F=1\rangle$

\qquad $|2\rangle=|5S_{1/2}, F=2\rangle$

\qquad $|3\rangle=|5P_{1/2}, F=2\rangle$

- Pump beam:

\qquad Wavelength $\lambda_p$ = 795 nm

\qquad Detuning to $|1\rangle \rightarrow |3\rangle$, $\Delta_p=\Delta_{12}=2\pi\times6.8~\rm GHz$

\qquad Polarization $\sigma^+$

\qquad $e^{-2}$-intensity waist diameter 1.6 mm

- Coupling beam:

\qquad Wavelength $\lambda_c$ = 795 nm

\qquad On resonance with $|2\rangle \rightarrow |3\rangle$

\qquad Polarization $\sigma^+$

\qquad Power 2.3 mW

\qquad $e^{-2}$-intensity waist diameter 2.3 mm

\qquad Rabi frequency $\Omega_c=2\pi\times14.5$ MHz

- Optical depth OD = 150

- Time bin width $\Delta t_{bin}$ = 10 ns

- For good EIT in Fig. 3(c) and (d) in the main text

\qquad Pump power 150 mW

\qquad Pump Rabi frequency $\Omega_p=2\pi\times218.6$ MHz

\qquad Detection time $T_c$ = 20 min

- For bad EIT in Fig. 3(e) and (f) in the main text

\qquad Pump power 100 mW

\qquad Pump Rabi frequency $\Omega_p=2\pi\times178.5$ MHz

\qquad Detection time $T_c$ = 120 min

\section{S2. Theory of degenerate biphoton generation}
\noindent\textbf{\textit{General Formalism --}} In this section, we present the formulas used for computing the theoretical plots in the main manuscript. By taking into account the spatial variations in the beam profiles of both the pump and coupling beams, we expand on the theory presented in Ref. ~\cite{SFWM_Du:08}. We express the two-photon relative wave function as
\begin{eqnarray}
\psi(\tau) &=& \frac{1}{2\pi}\int d\omega e^{-i\omega\tau} \int_{-L/2}^{L/2} dz \left[\kappa(\omega,z)e^{-i(k_c-k_p)z\cos{\theta}} \right. \nonumber \\
&&\times \left. e^{i\int_z^{L/2}dz'k_1(\omega,z')}e^{i\int_{-L/2}^{z}dz'k_2(\omega,z')}\right],
\label{eq:BiphotonWaveFunction1}
\end{eqnarray}
where $\tau=t_{1}-t_{2}$ represents the relative time delay, $L$ denotes the medium length, and $\omega=\omega_{1}-\omega_{0}$ is the frequency detuning of Photon 1. $k_c=\omega_c/c$ and $k_p=\omega_p/c$ are the wave numbers of the coupling and pump laser fields, respectively. In our SFWM system, characterized by time-reversal symmetry in the generation of paired photons from a single atom, the nonlinear parametric coupling coefficient is expressed as
\begin{eqnarray}
\kappa(\omega,z) = -i\frac{\omega_0}{2c}E_p(z)E_c(z) \left[\chi^{(3)}(\omega,z)+\chi^{(3)}(-\omega,z)\right], \nonumber \\
\label{eq:Kappa}
\end{eqnarray}
where $E_p(z)$ and $E_c(z)$ are the spatially varying electric field amplitudes of the pump and coupling laser fields, respectively. The third-order nonlinear susceptibility, \(\chi^{(3)}(\omega,z)\), is given by \cite{SFWM_Du:08}
\begin{eqnarray}
\chi^{(3)}(\omega,z)=\frac{N\mu_{13}\mu_{32}\mu_{24}\mu_{41}/(\varepsilon_0\hbar^3)}{(\Delta_{p}+i\gamma_{14})\left[|\Omega_c(z)|^2-4(\omega+i\gamma_{13})(\omega+i\gamma_{12})\right]},\nonumber \\
\label{eq:Chi3}
\end{eqnarray}
where $\mu_{ij}$ denote the electric dipole matrix elements, $\Omega_c(z)=\mu_{23}E_c(z)/\hbar$ is the coupling Rabi frequency, and $\gamma_{ij}$ represent dephasing rates. $\Delta_p=\omega_p-\omega_{31}$ is the pump detuning from the atomic transition $|1\rangle\rightarrow|3\rangle$. In the degenerate case with symmetry, we have
\begin{eqnarray}
k_1(\omega,z)=k_2(-\omega,z)=\frac{\omega_0+\omega}{c}\sqrt{1+\chi(\omega,z)},
\label{eq:k1}
\end{eqnarray}
where the linear susceptibility, $\chi(\omega,z)$, is defined as
\begin{eqnarray}
\chi(\omega,z) &=& \frac{4N|\mu_{13}|^2(\omega+i\gamma_{12})/(\varepsilon_0\hbar)}{|\Omega_c(z)|^2-4(\omega+i\gamma_{13})(\omega+i\gamma_{12})} \nonumber \\
&& + \frac{-N|\mu_{23}|^2(\omega+i\gamma_{13})/(\varepsilon_0\hbar)}{|\Omega_c(z)|^2-4(\omega+i\gamma_{13})(\omega+i\gamma_{12})}\frac{|\Omega_p(z)|^2}{\Delta_p^2+\gamma_{13}^2} \nonumber\\
&\approx& \frac{4N|\mu_{13}|^2(\omega+i\gamma_{12})
/(\varepsilon_0\hbar)}{|\Omega_c(z)|^2-4(\omega+i\gamma_{13}) (\omega+i\gamma_{12})},
\label{eq:chis}
\end{eqnarray}
considering the weak excitation condition $|\Omega_p(z)| \ll \Delta_p$.

\noindent\textbf{\textit{Special Case --}} For scenarios where the pump and coupling beams are uniform, Eq.~(\ref{eq:BiphotonWaveFunction1}) simplifies to
\begin{equation}
\psi(\tau)=\frac{L}{2\pi}\int d\omega \kappa(\omega)\Phi(\omega)e^{-i\omega\tau},
\label{eq:psi2}
\end{equation}
where $\Phi(\omega)$ represents the longitudinal detuning function, defined as
\begin{equation}
\Phi(\omega)=\text{sinc}\left(\frac{\Delta kL}{2}\right)e^{i(k_1+k_2)L/2}.
\end{equation}
Here, the complex phase mismatch is
\begin{eqnarray}
\Delta k(\omega)%&=&(\vec{k}_1+\vec{k}_2-\vec{k}_p-\vec{k}_c)\cdot \hat{z} \nonumber \\
= k_1(\omega)-k_2(\omega)-(k_c-k_p)\cos{\theta}.
\label{eq:phasemismatching}
\end{eqnarray}

In the EIT group delay regime, where the biphoton bandwidth is predominantly determined by the phase matching condition, $\kappa$ is approximately treated as a constant, $\kappa_0$, and $k_{1}(\omega)=k_2(-\omega)\simeq k_{0}+\omega/V_g+i\alpha$. The phase mismatch in Eq.~(\ref{eq:phasemismatching}) thus can be exprssed as $\Delta k \simeq 2\omega/V_g+\Delta k_{pc}$, where $\Delta k_{pc}=(k_p-k_c)\cos{\theta}$ . Consequently, we obtain
\begin{equation}
\Phi(\omega)=\text{sinc}\left(\frac{\omega L}{V_g}+\frac{\Delta k_{cp}L}{2}\right)e^{-\alpha L} e^{-ik_0L}.
\end{equation}
Its Fourier transform yields the biphoton wave function from Eq. (\ref{eq:psi2})
\begin{equation}
\psi(\tau)=\kappa_0 L\sqcap\left(\tau;-\frac{L}{V_g},\frac{L}{V_g}\right)e^{-i\frac{\Delta k_{cp}}{2}V_g\tau}e^{-\alpha L} e^{-ik_0L}.
\label{Biphotonwavefunction}
\end{equation}
Here, the rectangular function $\sqcap=1$ for $-\frac{L}{V_g}\leq \tau \leq \frac{L}{V_g}$, and $\Pi=0$ otherwise.

The biphoton coincidence counts are calculated as
\begin{equation}
CC(\tau)=\eta_d \eta_c|\psi(\tau)|^2\Delta t_{bin}T_c,
\end{equation}
where $\eta_d$ denotes duty cycle and  $\eta_c$ denotes the joint two-photon detection efficiency, $\Delta t_{bin}$ is the time bin width, and $T_c$ represents the collection time.

\section{S3. Time-Resolved Two-Photon Quantum Interference}
\noindent\textbf{\textit{Perfect Beam Splitter --}} The interference of two independent single photons incident on a beam splitter (BS) has been analyzed in Ref. \cite{APB.77.797}. In our SFWM biphoton source, the two photons exhibit energy-time entanglement. The interference of such photons on a BS is derived below. To observe temporal beating, we shift the optical frequency of the photons in channel 1 by $\delta$ using a pair of Acousto-Optic Modulators (AMOs) before they are incident on the BS, as shown in Fig. 4(a). Assuming a perfect 50\%:50\% BS, the transformation relations are $\hat{a}_3=\frac{1}{\sqrt{2}}(\hat{a}_2-\hat{a}_1)$ and $\hat{a}_4=\frac{1}{\sqrt{2}}(\hat{a}_2+\hat{a}_1)$. Without any time delay between paths 1 and 2, the two-photon wave packet at the detectors is given by
\begin{equation}
\begin{aligned}
\Psi_{3,4}(t_3, t_4)&=\langle0|\hat{a}_4(t_4)\hat{a}_3(t_3)|\Psi\rangle\\
&=\frac{1}{2}\langle0|[\hat{a}_2(t_4)+\hat{a}_1(t_4)][\hat{a}_2(t_3)-\hat{a}_1(t_3)]|\Psi\rangle\\
&=\frac{1}{2}[\Psi_{1,2}(t_3, t_4)-\Psi_{2,1}(t_3, t_4)],
\end{aligned}
\label{eq:Psi34}
\end{equation}
where
\begin{equation}
\Psi_{1,2}(t_3, t_4)=\langle0|\hat{a}_1(t_4)\hat{a}_2(t_3)|\Psi\rangle= e^{-i2\pi \delta t_4}\psi_0(t_4-t_3),
\label{eq:Psi12}
\end{equation}
\begin{equation}
\Psi_{2,1}(t_3, t_4)=\langle0|\hat{a}_2(t_4)\hat{a}_1(t_3)|\Psi\rangle= e^{-i2\pi \delta t_3}\psi_0(t_3-t_4).
\label{eq:Psi21}
\end{equation}
Defining $\tau=t_4-t_3$ and utilizing the exchange symmetry $\psi_0(\tau)=\psi_0(-\tau)$, we derive from Eqs. (\ref{eq:Psi34})-(\ref{eq:Psi21}):
\begin{eqnarray}
\Psi_{3,4}(t_3, t_4) 
=-ie^{-i\pi \delta (t_3+t_4)}\sin{(\pi\delta \tau)} \psi_0(\tau).
\end{eqnarray}
The two-photon Glauber correlation function is
\begin{eqnarray}
G_{34}(\tau)=|\Psi_{3,4}(t_3, t_4)|^2 
=\frac{1}{2}[1-\cos{(2\pi\delta\tau)}]|\psi_0(\tau)|^2, \nonumber\\
\end{eqnarray}
which exhibits temporal beating at frequency $\delta$ under the envelope $|\psi_0(\tau)|^2$. This beating, resulting from the two-photon interference, demonstrates the phase coherence of the biphoton wave packet.

\noindent\textbf{\textit{Imperfect Beam Splitter --}} In the case of an imperfect BS with reflectance $R$ and transmittance $1-R$, the transformation relations become $\hat{a}_3=\sqrt{R}\hat{a}_2-\sqrt{1-R}\hat{a}_1$ and $\hat{a}_4=\sqrt{1-R}\hat{a}_2+\sqrt{R}\hat{a}_1$. The wave packet at the detectors is then
\begin{equation}
\begin{aligned}
\Psi_{3,4}(t_3, t_4)&=R\Psi_{1,2}(t_3, t_4)-(1-R)\Psi_{2,1}(t_3, t_4)\\
&=e^{-i\pi \delta (t_3+t_4)}[Re^{-i\pi\delta\tau}-(1-R)e^{i
\pi\delta\tau}]\psi_0(\tau).
\end{aligned}
\end{equation}
The corresponding Glauber correlation function becomes
\begin{eqnarray}
G_{34}(\tau)&=&|\Psi_{3,4}(t_3, t_4)|^2 \nonumber\\
&=&[R^2+(1-R)^2-2R(1-R)\cos{(2\pi\delta\tau)}]|\psi_0(\tau)|^2. \nonumber\\
\label{eq:G34}
\end{eqnarray}
This modification leads to a degraded beating visibility, defined as
\begin{eqnarray}
V_0=\frac{2R(1-R)}{R^2+(1-R)^2},
\end{eqnarray}
which achieves its maximum value when $R=0.5$.
Accounting for the background accidental coincidence rate $n$, the visibility is adjusted to
\begin{eqnarray}
V&=&\frac{4R(1-R)|\psi_0(\tau)|^2}{2[R^2+(1-R)^2]|\psi_0(\tau)|^2+2n}\nonumber\\
&=&\frac{1}{\frac{1}{V_0}+\frac{2CC_n}{CC_{max}-CC_{min}}},
\end{eqnarray}

where $CC_n$ denotes the noise coincidence counts, and $CC_{max}$ ($CC_{min}$) represents the maximum (minimum)  of biphoton coincident counts, respectively.

At $\delta=0$, Eq.~(\ref{eq:G34}) simplifies to:
\begin{eqnarray}
G_{34,0}(\tau)=(2R-1)^2|\psi_0(\tau)|^2,
\label{eq:G341}
\end{eqnarray}
demonstrating a perfect Hong-Ou-Mandel interference $G_{34}(\tau)=0$ with a perfect BS ($R=0.5$) \cite{TPI_PhysRevLett.59.2044}. In our experiment, with $R=70\%$, this results in a nonzero residual coincidence
\begin{eqnarray}
G_{34,0}(\tau)=0.16\times|\psi_0(\tau)|^2,
\label{eq:G341_2}
\end{eqnarray}
as confirmed by our experimental measurements, shown in Fig. 4(b).

\bibliography{biphoton}